\documentclass[preprint,11pt,authoryear]{elsarticle}
\usepackage{amssymb}
\usepackage{lipsum}

\journal{Icarus}
\usepackage{lineno}
\usepackage{makecell}
\usepackage[a4paper, left=3.5cm, right=3.5cm, top=3.5cm, bottom=3.5cm]{geometry}
\usepackage{hyperref}
\usepackage{academicons}
\usepackage{xcolor}
\usepackage{url}
\usepackage{amsmath}
\usepackage{graphicx}
\usepackage{tabularx}
\usepackage{multirow}
\usepackage{booktabs}
\usepackage{siunitx}

\begin{document}
\begin{frontmatter}
\title{Quantitative Evaluation of the delta-Eddington, Hapke, and Shkuratov Models for Predicting the Albedo and Inferring the Grain Radius of Ice}

\author[first]{Aditya R. Khuller}
\affiliation[first]{organization={Polar Science Center, Applied Physics Laboratory},
            addressline={University of Washington}, 
            city={Seattle},
            state={WA},
            country={USA}}
\ead{akhuller@uw.edu}
\author[second]{Al Emran}
\affiliation[second]{organization={NASA Jet Propulsion Laboratory},
            addressline={California Institute of Technology}, 
            city={Pasadena},
            state={CA},
            country={USA}}

\begin{abstract}
Determining the physical properties of ices across the solar system is essential for understanding the surface dynamics, volatile transport, and climate evolution on ice-covered planetary bodies. Here, we use well-constrained measurements of snow that has metamorphosed into coarse-grained firn and bubbly glacier ice in East Antarctica to test three commonly-used radiative transfer models: delta-Eddington, Hapke, and Shkuratov. Using the measured optical properties, we find that the delta-Eddington model generally shows the least deviation from the measured albedo, followed by the Shkuratov and Hapke models, respectively. But when the models are used to infer the grain radius using the measured albedo, the Shkuratov model provides closer best-fit grain radii (off by average factor 0.9) than delta-Eddington (0.6), and Hapke (1.8). Despite this, the spectral albedos estimated by the Shkuratov and Hapke models using their respective best-fit grain radii deviate more from the measurements than delta-Eddington. This result is caused by the Hapke and Shkuratov models not accounting for: (1) the increased absorption within dense ice, and (2) specular reflection at the surface of firn and ice. Additionally, all three models do not account for the nonsphericity of bubbles within ice. The combination of these factors leads to model errors generally increasing with increasing grain radius. Based on our quantitative comparison, we recommend using the delta-Eddington model for predicting the albedo and inferring the grain radius of ices across the solar system because it generally produces the least error while using realistic physical parameters.
\end{abstract}

\end{frontmatter}

\section{Introduction}
\label{intro}
Accurately constraining physical properties of ices, such as abundance and grain radius, across the solar system is important for understanding the formation and evolution of the surfaces of planetary bodies. Moreover, the absorbed solar radiation is often the largest term in the surface energy budget. Predicting the albedo (the integral of the reflectance over all angles, i.e., the flux of upward radiation divided by the flux of downward radiation) of ices at the surface is therefore essential for understanding surface dynamics, volatile transport, and the evolution of climate. 

Over the last three decades, numerous authors \citep[e.g.,][]{buratti1985application, calvin1994spatial, Poulet2002, ciarniello2011hapke,protopapa2014water, gyalay2019nonlinear, emran2022uncertainty, pascuzzo2022sensitivity} have used approximate models of radiative transfer such as the Hapke model \citep{hapke1981bidirectional, hapke2001space, hapke2012theory} and the Shkuratov model \citep{shkuratov1999modeling} to infer the optical properties, and to predict the albedo of ices on the Moon, Mars, and beyond. Those models were originally developed for porous, non-ice media such as lunar regolith and soil \citep{hapke1981bidirectional, shkuratov2005regolith, shepard2007test}. Some limited laboratory studies have been performed to compare the reflectance of ices to those models, with mixed results. For example, \citet{yoldi2015vis} found that the grain radii required by the Hapke model (12 $\mu$m) to match the measured reflectance of laboratory ice were a factor of 16 larger than the actual grain radii (0.75 $\mu$m). However, small ice particles quickly sinter to form agglomerates under laboratory conditions, so constraining the actual grain radius can be challenging \citep[e.g.,][]{jost2016experimental}. Additionally, making laboratory measurements of ice albedo are challenging because ice is very weakly absorbing at visible/near-infrared wavelengths, requiring special designs for laboratory apparatus \citep{light2015albedo}. Thus, it is unclear whether those models can be extended to ice, because of the weak absorption of ice in comparison with the materials of planetary surface regolith.

This paper tests three commonly-used radiative-transfer models against spectral albedo (also known as spectral directional-hemispherical reflectance) measurements for surfaces with well-documented microstructure \citep{dadic2013effects}: Antarctic snow that has metamorphosed into coarser-grained firn (density 550--830 kg m$^{-3}$) and bubbly glacier ice (density 830--917 kg m$^{-3}$). Two of the models are those mentioned above: Hapke and Shkuratov. The third uses the delta-Eddington method \citep{joseph1976delta}, based on work validated previously using measurements of snow, firn, and glacier ice on Earth \citep{wiscombe1980ice, mullen1988theory, briegleb2007delta, dadic2013effects, dang2019intercomparison, flanner2021snicar, whicker2022snicar, khuller2024photosynthesis} and Mars \citep{khuller2021spectral}.

By performing a quantitative comparison of these three models, we aim to make recommendations for future analyses of ices in the solar system (e.g., which model to use depending on the given scenario, what errors to expect, and potential methods to account for errors).

\section{Methods}
\subsection{Spectral albedo measurements of firn and glacier ice in East Antarctica}
\label{sec2.1}
\begin{figure*}
	\centering 
	\includegraphics[width=1.\textwidth, angle=0]{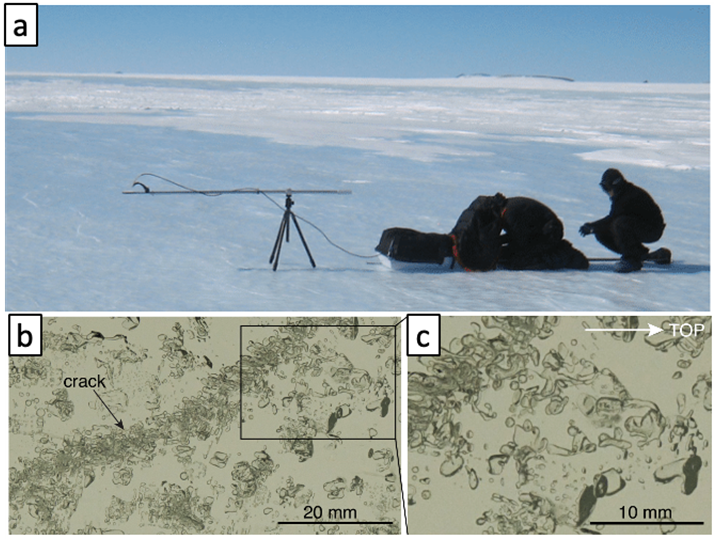}	
	\caption{(a) Spectral albedo measurement setup over ice near the Allan Hills in East Antarctica. (b) Thick ice section of glacier ice of density 856 kg $m^{-3}$ showing irregular bubbles and cracks. (c) Zoomed-in section of black box shown in (b). Images adapted from \cite{dadic2013effects}.} 
	\label{fig1}%
\end{figure*}

Dadic et al. \citep{dadic2013effects} measured the spectral albedo and collected core samples of firn and glacier ice near the Allan Hills in East Antarctica (76.67$^\circ$S, 159.23$^\circ$E). The spectral albedo was measured under uniform overcast clouds to ensure diffuse illumination (which ensures minimum measurement error), using a FieldSpec Pro JR spectral radiometer from Analytical Spectral Devices, Inc. The instrument measures radiance across wavelengths 0.35-2.5 $\mu$m at 0.003-0.01 $\mu$m spectral resolution; the same instrument was used previously to measure snow albedo, snow transmittance, and bidirectional reflectance at Dome C on the East Antarctic Plateau \citep{hudson2006spectral, warren2006light}. The albedo measurements were made by suspending a Spectralon diffuser-plate receptor on a rod 70 cm above the ice, supported by a tripod on the other end (Fig. \ref{fig1}a). The downwelling and upwelling radiation were measured alternately 10 times at each site. Then, the spectral flux and time for the five downwelling measurements were interpolated to obtain the downwelling radiation at the same time that the upwelling radiation was measured, before being averaged for each site. The spectrally averaged standard deviation was 0.005 and deemed negligible \citep{dadic2013effects}. Shadows were minimal under diffuse incidence, and accounted for by multiplying the albedo by 1.017 based on a geometric analysis for the same measurement setup \citep{brandt2011controlled}. The firn and ice material were characterized by micro-computed tomography, from which the density and the area/mass ratio could be computed.  

While there are many other measurements of spectral albedo of snow and ice on Earth \citep[e.g.,][]{obrien1975red, grenfell1994reflection, carmagnola2013snow}, we use the albedo and micro-computed tomography measurements of \citep{dadic2013effects} because most other measurements do not have tight constraints on key model inputs such as the grain radii, density, and impurities. Additionally, we prefer to use field measurements because laboratory measurements of snow and ice spectral albedo are challenging to make due to the samples' low optical absorption, and complications involved with shadowing, edge effects, and uniformity of illumination \citep[Section 1 of ][]{light2015albedo}. 

\begin{table}[htbp]
\centering
\small  
\caption{Characterization of firn and ice at three sites using micro-computed tomography \citep{dadic2013effects}. The albedos measured at these sites are used in our comparison of radiative transfer models.}
\begin{tabular}{l l c c c}
\hline
Name & Sample name & SSA (m$^2$ kg$^{-1}$) & radius $r$ (mm) & Density (kg m$^{-3}$) \\
\hline
Fine Firn & R9 Firn & 3.8 & 0.86 & 668 \\
Coarse Firn & R7 Firn & 1.9 & 1.72 & 777 \\
Ice & R1 Ice & 0.39 & 8.39 & 864 \\
\hline
\end{tabular}
\label{table1}
\end{table}
\subsection{Specific surface area measurements of firn and glacier ice in Eastern Antarctica}
\label{sec2.2}
The specific surface area (SSA) is the area of air-ice interfaces per unit mass of ice, which is reciprocally related to the effective grain radius ($r$):

\begin{equation}
\mathrm{SSA} = \frac{3}{r \, \rho_\mathrm{ice}},
\end{equation}
where $\rho_\mathrm{ice} = 917$ kg m$^{-3}$ is the density of pure water ice  \citep{cuffey2010physics}.  It is easier to accurately measure the SSA of snow and ice than measure the grain radius directly. For example, snow grains may not be spherical, so defining a radius can be challenging. In addition, individual grains are often difficult to distinguish, because they are in contact. To derive the SSA and density for use as model input \citep{dadic2013effects} collected core samples (1 m long, 90 mm diameter) of the firn and glacier ice at the locations of the spectral albedo measurements. Those samples were analyzed using micro-computed tomography to obtain the SSA and density  \citep[see Section 4.1 of][]{dadic2013effects}.  For our modeling comparison, we selected the data from three albedo-SSA measurements, shown in Table \ref{table1}.

\subsection{Radiative transfer models}
\label{sec2.3}
Using three radiative transfer models, we simulated the spectral albedo for each of the three measurements made by \cite{dadic2013effects}. In all cases, we used H$_2$O ice refractive indices from \cite{warren2008optical}.  For our first set of simulations (Section \ref{sec3.1}), we used the measured SSA (in the form of grain radii) and density values as inputs for the model, without adjusting any other parameters. For our second set of simulations (Section \ref{sec3.2}), we inferred the best-fit grain radius for each measurement by matching the observed albedo. Below we describe each model in more detail.

Note that we avoid using the ambiguous term ‘grain size’, and instead use grain radius throughout, to avoid confusion. Since the Hapke and Shkuratov models require ‘grain size’ as input, we have assumed that their grain size is equivalent to the grain diameter \citep{shkuratov1999modeling, Poulet2002}.

\subsubsection{The Hapke model}
\label{sec2.3.1}
There are many versions of the Hapke model \citep[e.g.,][]{hapke1981bidirectional, roush1994charon, hapke2001space, hapke2012theory}. In this work, we evaluate a commonly used version of the Hapke model presented by \cite{roush1994charon}, which assumes isotropically-scattering particles. Numerous authors \citep[e.g.,][]{davies1997detection, cruikshank1998composition, brown2008louth, seelos2008geomorphologic, becerra2015transient} have used this version of the Hapke model to study ices in the solar system.

According to \citet[][p. 289]{hapke2012theory}, the direct-beam albedo of a medium of isotropically scattering, well-separated particles using the two-stream approximation  \citep{schuster1905} to radiative transfer is given by:
\begin{equation}
A_\mathrm{Hapke}(\mu_0) = \frac{1 - \gamma}{1 + 2 \gamma \mu_0},
\end{equation}
where
$\gamma = \sqrt{1 - \bar{\omega}}$ \citep[][p. 191]{hapke2012theory},
$\bar{\omega}$ is the single-scattering albedo (the probability of scattering, $0 \leq \bar{\omega} \leq 1$), and $\mu_0$ is the cosine of the solar zenith (incident) angle.  $\bar{\omega}$ is approximated by Equation 18 of \cite{hapke2001space}:
\begin{equation}
\bar{\omega} = S_e + \frac{(1 - S_e)(1 - S_i) \Theta}{1 - S_i \Theta},
\end{equation}
where
\begin{align}
S_e &= \frac{(n - 1)^2 + k^2}{(n + 1)^2 + k^2} + 0.05 \quad \text{(Equation 21 of \citealp{warell2010hapke})} \\
S_i &= 1.014 - \frac{4}{n(n+1)^2} \quad \text{(Equation 6 of \citealp{lucey1998model})}
\end{align}are the Fresnel reflection coefficients for externally and internally incident light averaged over all incidence angles, respectively. Here, $n$ and $k$ are the real and imaginary parts of the complex refractive index, $m = n + i k$. The internal-transmission parameter $\Theta$ is calculated using Equation 21 of \cite{hapke2001space}, assuming the internal scattering coefficient is zero. Eqs. 5 to 8 of \cite{emran2023discrepancy} provide further details.  Note that some authors \citep[e.g.,][]{lucey1998model, lawrence2007radiative, warell2010hapke, li2011radiative} calculate the absorption coefficient as: 
\begin{equation}
\beta_\mathrm{abs} = \frac{4 \pi n k'}{\lambda}, \quad \text{where} \quad k' = \frac{k}{n},
\end{equation}
and referred to as the “coefficient of the imaginary index of refraction”, although it is sometimes mistaken for $k$. We instead use
\begin{equation}
\beta_\mathrm{abs} = \frac{4 \pi k}{\lambda},
\end{equation}
which is what is recommended by \citet[p. 88]{hapke2012theory}, and indeed most commonly used for the absorption coefficient \citep[e.g.,][p. 77]{petty2006first}.  

Under overcast skies (e.g., due to clouds or atmospheric aerosols), the illumination is diffuse rather than direct. The diffuse albedo 
$A_\mathrm{Hapke}$  can be obtained by assuming isotropically incident radiation and integrating Equation (2) over all angles of incidence:
\begin{equation}
A_\mathrm{Hapke} = 2 \int_0^1 \mu_0 \, A_\mathrm{Hapke}(\mu_0) \, d\mu_0.
\end{equation}

\subsubsection{The Shkuratov Model}
\label{sec2.3.2}
The second radiative transfer model commonly used to calculate the spectral albedo of planetary ices is the model of \cite{shkuratov1999modeling}. Unlike the Hapke model, the spectral albedo calculated using the Shkuratov model is independent of viewing geometry. Instead, the model results are dependent, for a given set of optical constants, on the grain diameter (which we have converted to radius for model intercomparisons) and porosity $\phi$ of the medium, i.e., $1-$ volume fraction of ice, in our case. The spectral albedo of an optically thick medium is given by Equation 12 of \cite{shkuratov1999modeling}:

\begin{equation}
A_{\text{Shkuratov}} = \frac{1 + p_b^2 - p_f^2}{2 p_b} - 
\sqrt{\left( \frac{1 + p_b^2 - p_f^2}{2 p_b} \right)^2 - 1}, 
\end{equation}

where

\begin{align}
p_f &= (1 - \phi) \, r_f + \phi, \\
p_b &= (1 - \phi) \, r_b,
\end{align}are the forward and backward components of the angularly-averaged indicatrix that governs the scattering behavior of particles \citep{shkuratov1999modeling}. $r_f$ and $r_b$ represent the fraction of light scattered into the forward and backward hemispheres, respectively  \citep[Equation 10 of][]{shkuratov1999modeling}.

\subsubsection{The delta-Eddington Model}
\label{sec2.3.3}

The delta-Eddington model was initially developed to calculate spectral fluxes within the atmosphere by modifying Eddington's solution to the radiative transfer equation \citep{joseph1976delta}. A few years later, it was successfully used to explain the spectral albedo of snow on Earth  \citep{warren1980ice,wiscombe1980ice}. Note that although some authors classify the Eddington approximation as a type of two-stream approximation, it is important to emphasize the difference between the two methods. The Eddington method assumes that the intensity is a linear function of the cosine of the angle from the vertical  (the zenith angle; \citealp{eddington1916radiative}). By contrast, the two-stream approximation assumes that the intensity is constant with angle in the upward hemisphere, with a different constant value in the downward hemisphere \citep{schuster1905}. 

The direct-beam albedo $A_{\text{delta-Eddington}}(\mu_0)$ of a single-layer using the delta-Eddington method is given by:

\begin{equation}
A_{\text{delta-Eddington}}(\mu_0) = A_n \exp\left(-\frac{\tau}{\mu_0}\right) + B_n \left[ \exp(\epsilon \tau) - \exp(-\epsilon \tau) \right] - K_n,
\end{equation}

where $A_n$, $B_n$, $K_n$, and $\epsilon$ are determined by $\bar{\omega}$, the asymmetry parameter $g$ (the mean value of the cosine of the scattering angle), the optical depth $\tau = z (\beta_{\text{abs}} + \beta_{\text{sca}}) = z \beta_{\text{ext}}$, and $\mu_0$ (see Equation set 50 of \citealp{briegleb2007delta}). Here, $z$ is the depth, and $\beta_{\text{abs}}$, $\beta_{\text{sca}}$, and $\beta_{\text{ext}}$ are the absorption, scattering, and extinction coefficients, respectively. A detailed description of the calculation of these optical properties is provided in Appendix A of \cite{khuller2021spectral}. 

We use a multi-layer delta-Eddington model \citep{khuller2024photosynthesis} that incorporates the method of \cite{whicker2022snicar} to allow for the inclusion of a specular refractive layer to account for refraction at boundaries between air, ice, and liquid water (Equations 7–12 of \citealp{whicker2022snicar}) that becomes particularly important for modeling the albedo of firn and glacier ice \citep{dadic2013effects}. As described above for the Hapke model, the albedo under diffuse illumination, $A_{\text{delta-Eddington}}$ is obtained by integrating over all incidence angles (Eq. 8).

\section{Results}
\label{sec3}
\begin{figure*}
	\centering 
	\includegraphics[width=1.\textwidth, angle=0]{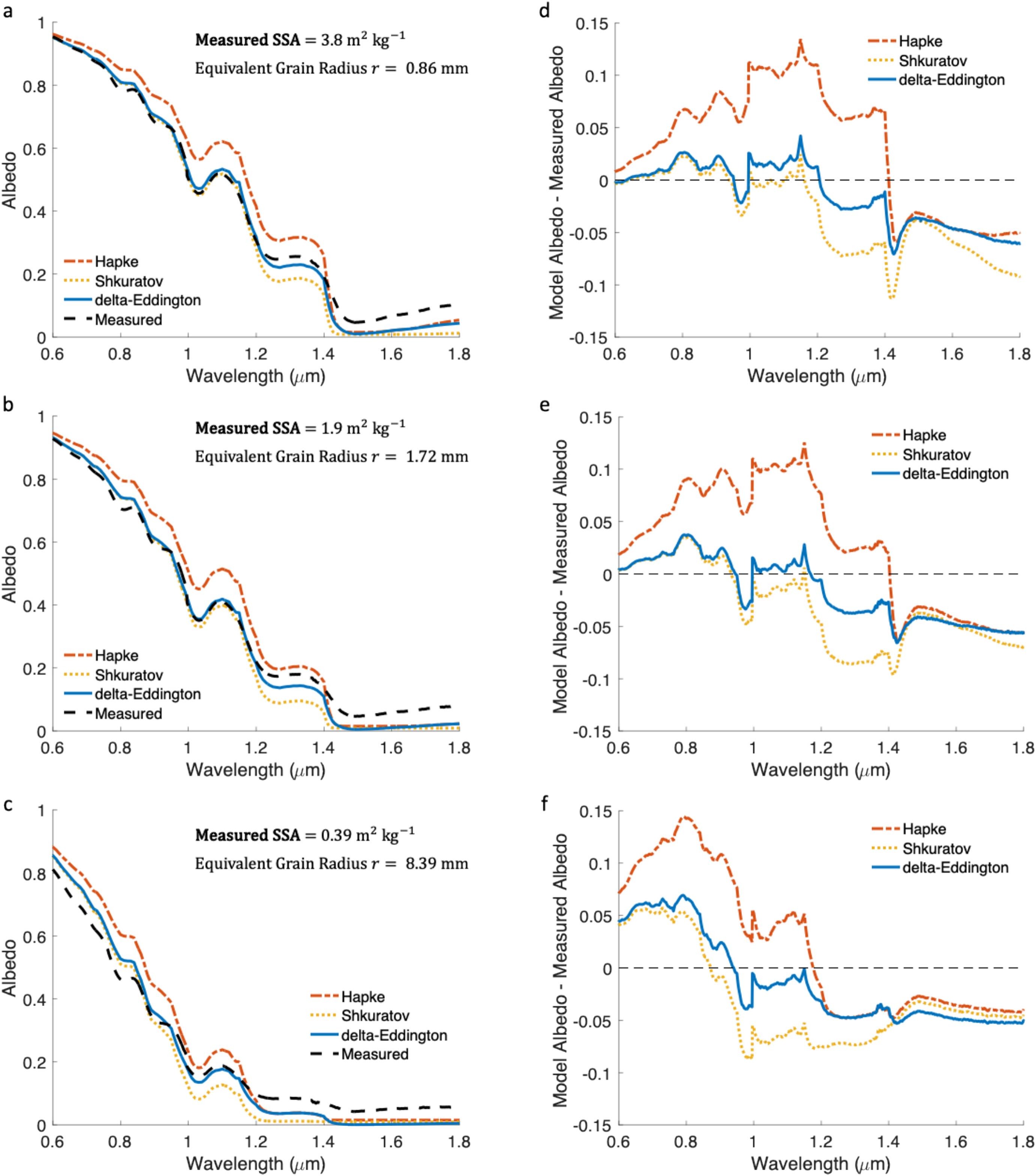}	
	\caption{((a–c) Comparison of calculated spectral albedo with \citet{dadic2013effects} measurements of firn and glacier ice in East Antarctica under diffuse illumination. The specific surface area (SSA $= 3/(r \rho_{\rm ice})$) is related to an effective grain radius $r$ (Eq.1). The SSA was measured using micro-computed tomography \citep{dadic2013effects} and used for the calculations.  (d–f) Difference between model and measured spectral albedo.} 
	\label{fig2}%
\end{figure*}

\begin{figure*}
	\centering 
	\includegraphics[width=1.\textwidth, angle=0]{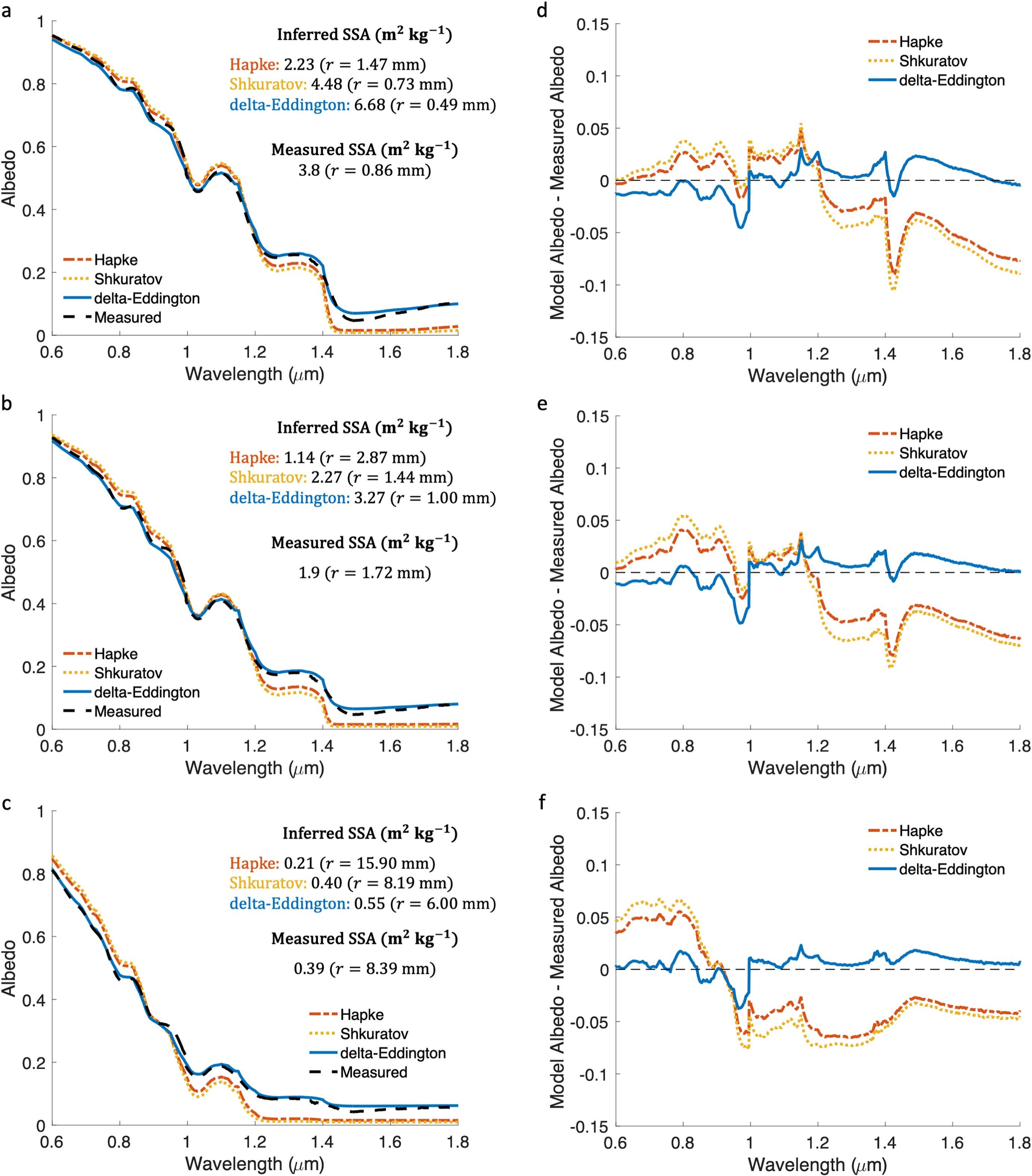}	
	\caption{(a–c) Comparison of calculated spectral albedo using best-fit parameters for each model with \citet{dadic2013effects} measurements of firn and glacier ice in East Antarctica under diffuse illumination.  (d–f) Difference between model and measured albedo.  The delta-Eddington model we use \citep{khuller2024photosynthesis} accounts for refractive boundaries often found in firn and bubbly glacier ice, which can cause an increase in albedo around $1.5~\mu$m due to specular reflection, where the multiple-scattering albedo is negligible \citep{mullen1988theory, briegleb2007delta, dang2019intercomparison, whicker2022snicar}.} 
	\label{fig3}%
\end{figure*}

\begin{figure*}
	\centering 
	\includegraphics[width=1.\textwidth, angle=0]{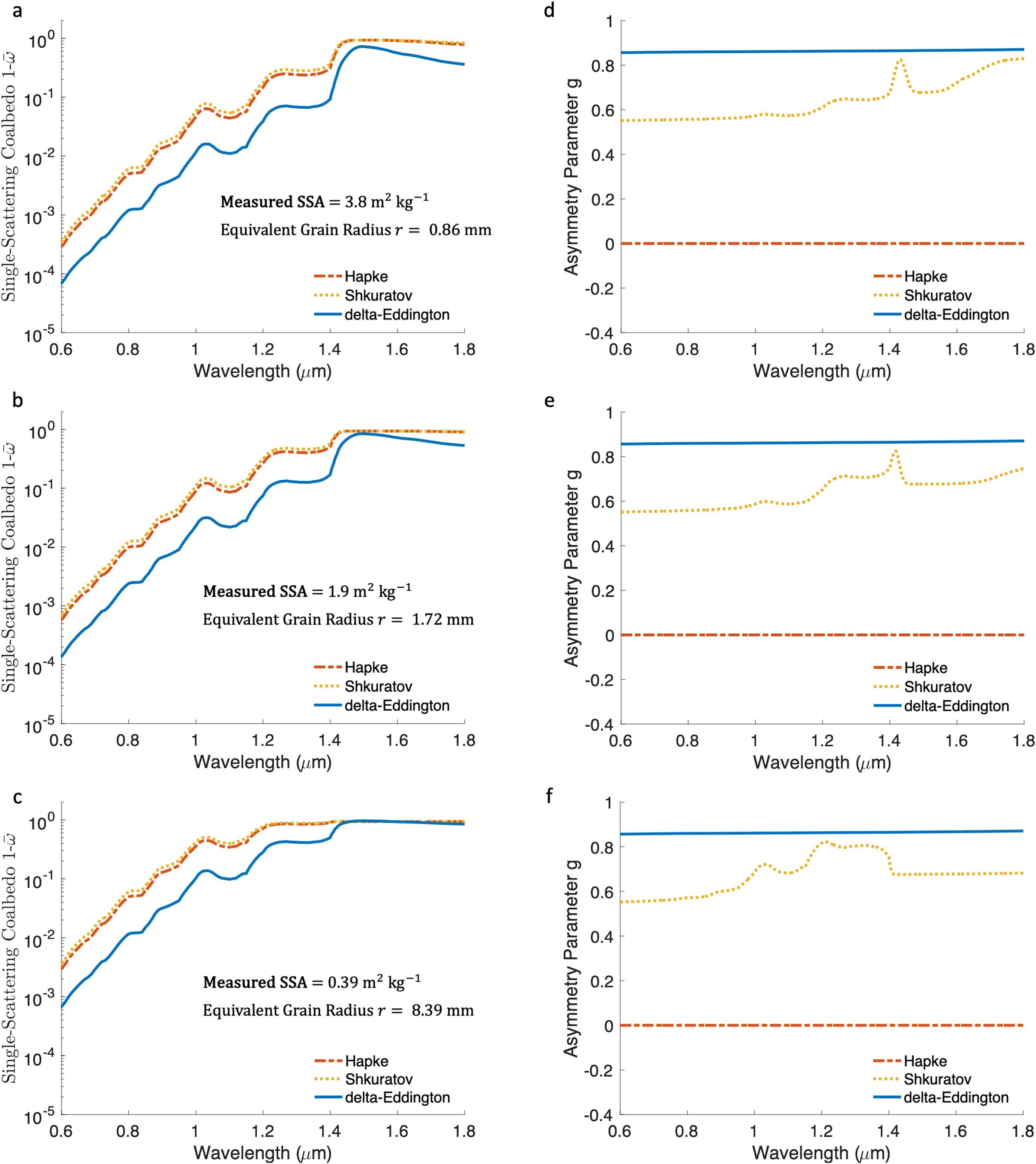}	
	\caption{(a–c) Comparison of the single-scattering co-albedo $(1-\bar{\omega})$ computed by each model using \citet{dadic2013effects} measurements of firn and glacier ice SSA and density in East Antarctica.  (d–f) The asymmetry parameter $g$ for each model is shown. For the Shkuratov model, \citet{Poulet2002} show that these single-scattering parameters can be derived using:  
	$\bar{\omega}_{\rm Shkuratov} = r_b + r_f, \quad 
	g_{\rm Shkuratov} = \frac{r_f - r_b}{r_f + r_b}$.} 
	\label{fig4}%
\end{figure*}

\subsection{Model comparison using measured specific surface area}
\label{sec3.1}

Figure \ref{fig2} shows the spectral albedo calculated using the three models, compared to the measurements of \citet{dadic2013effects}. As model input for each case, we used the grain radius derived using the SSA measurements. For this comparison, we did not vary any model inputs to attempt to match the measurements; for example, we did not use a specular refractive layer in our delta-Eddington model. We compared the model results with the measurements only for wavelengths 0.6–1.8~$\mu$m (black dots) because wavelengths short of  0.6~$\mu$m were affected by absorption by trace amounts of volcanic ash (see Figure 15 of \citealp{dadic2013effects}), and the data were noisy beyond 
1.8~$\mu$m because of strong absorption by atmospheric water vapor \citep{dadic2013effects}.  

In all cases, the Hapke model predicts the highest albedo at essentially all wavelengths, followed by the delta-Eddington and Shkuratov models, respectively. None of the three models match the measurements well for $\lambda>1.2~\mu$m. At visible wavelengths ($\lambda \le 0.7~\mu$m), the delta-Eddington and Shkuratov models match the firn albedo well (Figs. \ref{fig2}a and \ref{fig2}b) but overestimate the albedo for ice (Fig. \ref{fig2}c).  

Overall, the delta-Eddington model produces the lowest difference between the modeled and measured spectra, although the Shkuratov model produces similarly low albedo differences for $\lambda<0.8 \mu$m. The difference between model and measured albedo increases with grain radius in all cases. Beyond 1.4 $\mu$m, the multiple-scattering albedo of firn and ice is so low that specular reflection causes the albedo to be greater than the model albedos by 0.05.

\subsection{Inferring Grain Radii using Best-Fit Parameters for each Model}
\label{sec3.2}
After comparing the model results using the measured SSA and density as inputs, we sought to compare the best-fit grain radius inferred using each model, given the measured albedo. Thus, for each model, we varied the grain radius to give the best fit to the albedo measurements. Since the delta-Eddington model we are using was designed for modeling firn and glacier ice that can have a specular refractive boundary at the surface, we also tested whether the use of a refractive boundary in the delta-Eddington model improved the matches.  

Figure~\ref{fig3} shows the results of our comparison of modeled albedos estimated from best-fit radii using the three models in comparison with the albedo measurements of \citet{dadic2013effects}. Incorporating a specular refractive boundary at the surface in the delta-Eddington model gave the best spectral albedo results, as shown previously by \citet{dadic2013effects}. Apart from $\lambda=1.5 \mu$m (where specular reflection dominates the albedo), specular reflection reduces the albedo because upward radiation is reflected back into the ice, giving photons another chance to be absorbed  \citep{mullen1988theory}. Using the best-fit radii for each case (Fig. \ref{fig3}) instead of the measured radii (Fig. \ref{fig2}) improves the fit to the albedo measurements considerably, especially for the Hapke model at wavelengths short of 1 $\mu$m.  In contrast, the Shkuratov model errors increase for $\lambda<1 \mu$m, but reduce overall at longer wavelengths in comparison with the measured radii result (Fig. \ref{fig2}). The delta-Eddington model error reduces significantly overall, to $<0.05$ in all cases.  

For the Hapke and Shkuratov models, the albedo error still increases with grain radius, because they were not designed to model firn and glacier ice. In contrast to those two models, the delta-Eddington model performs better with increasing grain radius.  

In all cases, the Hapke model overestimates the grain radii, whereas the Shkuratov and delta-Eddington models underestimate the grain radii (Fig. \ref{fig3}). Overall, the Shkuratov model provides the closest best-fit radii, despite having larger errors in modeled spectral albedo than the delta-Eddington model. For example, for the 0.86 mm firn, the Hapke model overestimates the grain radius by a factor of 1.7, whereas the Shkuratov, and delta-Eddington models underestimate the grain radius by factors of 0.8, and 0.6, respectively (a factor of 1 would indicate perfect agreement between model and measurement). For the 1.72 mm firn, the models are off by similar factors of 1.7, 0.8, and 0.6, respectively. Finally, for the ice, while the Hapke model result worsens, the Shkuratov model matches the data, and the delta-Eddington result improves slightly, so that the models are off by factors of 1.9, 1 (the Shkuratov model matches the true grain radius), and 0.7, respectively.

\subsection{Errors in Computing Broadband (spectrally-weighted) Albedos}
\label{sec3.3}
While the spectral albedo comparisons are important for understanding the effects of changing grain radii, estimating the broadband (spectrally-weighted) albedo is what is needed for the energy budget of ice-covered surfaces. The spectrally-weighted albedo $\bar{A}$ is obtained by weighting the spectral albedo $A(\lambda)$ by the downward solar spectrum $S(\lambda)$:

\begin{equation}
\bar{A} = \frac{\int_{0.6~\mu\rm m}^{1.8~\mu\rm m} S(\lambda) A(\lambda) d\lambda}{\int_{0.6~\mu\rm m}^{1.8~\mu\rm m} S(\lambda) d\lambda}.
\end{equation}

For each calculation, we used the cloudy sky downward solar spectrum 
$S(\lambda)$ for atmospheric conditions on the Antarctic Plateau \citep[Figure 1 of][]{brandt1993solar}. As described above, we only used wavelengths 0.6–1.8~$\mu$m to avoid contamination by volcanic tephra and to exclude instrument noise.  

Table~\ref{table2} shows the difference between the measured spectrally-weighted albedo and the modeled spectrally-weighted albedo using the measured grain radii and the best-fit grain radii for each case. In all cases, the errors (difference between measured and modeled) in calculating the spectrally-weighted albedo are less than 0.05, except for the Hapke model result for coarse firn and ice, using the measured grain radius. Note that while these errors may seem small, errors of this magnitude can be important for determining ice loss rates from melting and/or sublimation. For example, an error in albedo of 0.05 leads to an error of 20 W m$^{-2}$ in absorbed solar radiation $F_{\rm abs}$ for a typical Arctic summertime daily incident solar flux $F_{\rm down}$ of 400 W m$^{-2}$,  

\begin{equation}
F_{\rm abs} = F_{\rm down} (1-Albedo) \, ,
\end{equation}

\noindent as noted by \citet{warren2019light}. By comparison, the radiative forcing for doubling the amount of CO$_2$ in Earth’s atmosphere is just 4 W m$^{-2}$ \citep[e.g.,][]{hansen1997radiative}.  

When using the measured grain radii as input for each model, the Hapke model and the delta-Eddington model underestimate the broadband albedo, whereas the Shkuratov model overestimates the broadband albedo, except for the ice case. The Shkuratov and delta-Eddington model errors are lower than the Hapke model errors by an order of magnitude.  

When the models are used to infer the best-fit grain radius, the albedo errors generally decrease by an order of magnitude for the Hapke model. This reduction in Hapke model error is caused by it fitting the measurements well at wavelengths near 0.6 $\mu$m, close to where solar radiation peaks. On the other hand, the Shkuratov model albedo errors increase when using the best-fit radius. This is due to better albedo fits overall, but poorer fits at wavelengths near 0.6 $\mu$m. The delta-Eddington model albedo error increases for fine firn, but decreases for coarse firn and ice.

\begin{table*}[htbp]
\centering
\caption{Comparison of spectrally-weighted error in albedo using measured and best-fit inferred grain radii for each model in comparison with \cite{dadic2013effects} observations. Values represent the difference between measured broadband albedo and model broadband albedo.}
\sisetup{table-format=-1.4, round-mode=places, round-precision=4}
\begin{tabular}{l l S S S}
\toprule
Grain Radius &  & \multicolumn{3}{c}{Measured - Model Broadband Albedo} \\
\cmidrule(lr){3-5}
 & & {Hapke} & {Shkuratov} & {delta-Eddington} \\
\midrule
\multirow{3}{*}{Measured} 
 & Fine Firn & -0.0496 & \texttt{+}0.0053 & -0.0042 \\
 & Coarse Firn & -0.0574 & \texttt{+}0.0031 & -0.0063 \\
 & Ice & -0.0755 & -0.0036 & -0.0238 \\
\midrule
\multirow{3}{*}{Best-fit Inferred} 
 & Fine Firn & -0.0049 & -0.0101 & \texttt{+}0.0074 \\
 & Coarse Firn & -0.0089 & -0.0149 & \texttt{+}0.0047 \\
 & Ice & -0.0093 & -0.0125 & -0.0027 \\
\bottomrule
\end{tabular}
\label{table2}
\end{table*}


\section{Implications for Predicting the Albedo and Inferring the Grain Radius of Ices across the Solar System}
\label{sec4}

The results in Section \ref{sec3} show that no model is able to completely match the observed spectral albedo using the measured SSA (or equivalently, grain radius). The errors generally increase with decreasing SSA (or increasing grain radius). Although the errors vary with wavelength and SSA, the delta-Eddington model generally shows the least deviation from the measured albedo, followed by the Shkuratov and Hapke models, respectively.  

When the models are used to infer the grain radius, the Shkuratov model provides the closest best-fit grain radii overall (off by average factor of 0.9), followed by the delta-Eddington (factor of 0.6) and Hapke (factor of 1.8) models, which either underestimate or overestimate the grain radii. However, despite providing the closest best-fit grain radii, the Shkuratov model (and similarly, the Hapke model) provides relatively poor fits to the measured albedo. In contrast, the delta-Eddington model provides excellent fits to the measured albedo whilst underestimating the grain radii. The fact that the Hapke and Shkuratov models provide closer best-fit grain radii while outputting spectral albedos that deviate significantly from the measurements suggests those models are lacking the ability to capture some key physical processes, as we shall discuss in more detail below.  

Errors in calculating the spectral albedo also leads to errors in calculating the broadband (spectrally-weighted) albedo. The Shkuratov model produces the lowest errors in broadband albedo using the measured grain radii, but the results are mixed using the inferred best-fit grain radii. For fine firn, the Hapke model produces the lowest error in broadband albedo, whereas the delta-Eddington model performs the best for coarse firn and ice.  

The models fail to match the measurements for three main reasons. The first reason applies to the Hapke and Shkuratov models, which were designed to simulate porous media with a non-absorbing medium such as a vacuum, or air between the grains \citep[e.g.,][]{hapke1981bidirectional, shkuratov1999modeling, hapke2001space}. But firn and glacier ice are composed of  $>60\%$ ice, o they can be considered absorbing media, particularly for $\lambda > 1 \mu\mathrm{m}$, where ice is more strongly absorbing \citep{mullen1988theory,dadic2013effects}. Thus, while these models perform adequately for small grain radii, their errors increase with increasing grain radius. Note that the commonly-used version of the Hapke model we apply does not explicitly account for porosity, which could potentially improve the Hapke model results. Nevertheless, porosity has no effect on albedo per se since it is the SSA of ice that determines the albedo of snow, firn, and ice \citep{bohren1979snowpack}. The Hapke model also has numerous other parameters that can be potentially varied to improve the model results, but the vast number of unconstrained parameters can make them difficult to constrain \citep{shkuratov2012critical,hapke2013comment}.  

The second reason is based on how each model accounts for single-scattering. Figure \ref{fig4} shows how the single-scattering coalbedo $(1-\bar{\omega})$  and the asymmetry parameter $g$ vary as a function of wavelength using the \cite{dadic2013effects} measurements of SSA and density as input for each model. Note that although the Shkuratov model does not explicitly define $\bar{\omega}$ and $g$, \citet{Poulet2002} show that these single-scattering parameters can be defined using the Shkuratov model parameters $r_f$ and $r_b$:  
\begin{equation}
\bar{\omega}_{\rm shkuratov} = r_b + r_f, \quad \text{and} \quad
g_{\rm shkuratov} = \frac{r_f - r_b}{r_f + r_b}.
\end{equation}

In all cases, $(1-\bar{\omega})$  increases with increasing grain radius, because the likelihood of absorption increases with grain radius (Figs. \ref{fig4}a-c). Figure \ref{fig4} also shows that $(1-\bar{\omega})_{\rm delta\text{-}Eddington}$ is significantly lower than $(1-\bar{\omega})_{\rm Hapke} \simeq (1-\bar{\omega})_{\rm Shkuratov}$ for $\lambda < 1 ~\mu\mathrm{m}$, and all the models are similar for $\lambda > 1.4 ~\mu\mathrm{m}$, particularly for the ice case. The fact that $(1-\bar{\omega})_{\rm delta\text{-}Eddington}$ is lower than the other single-scattering coalbedo of the other two models would imply that the albedo predicted by the delta-Eddington model would be the highest, all else being equal.  

The asymmetry parameter $g$ is distinct for each model, with $g_{\rm Hapke} = 0$  (assuming isotropic scattering, as is commonly assumed for analysis of planetary ices, unless g is well constrained by measurements of the phase function, which are rare), $g$ is well-constrained by measurements of the phase function, which are rare), $0.55 \le g_{\rm Shkuratov} \le 0.83$, and $g_{\rm delta\text{-}Eddington}$ increases slightly from 0.86 at $\lambda = 0.6~\mu\mathrm{m}$ to 0.87 at $\lambda = 1.8~\mu\mathrm{m}$, assuming scattering by air bubbles within ice  \citep{mullen1988theory}. Note that this asymmetry parameter $g$ is commonly listed as $\xi$ in the Hapke model \citep[p. 71]{hapke2012theory}.

The albedo is critically dependent on the value of $g$ because mean photon lengths increase with increasing $g$, and large $g$ concentrates the scattering angles toward the forward direction, thereby increasing the probability of photon absorption, and reducing the albedo. Thus, all else being equal, an increase in $g$ results in a decrease in albedo \citep[Figure 19 of ][]{wiscombe1980ice}. Since $g_{\rm Hapke}=0$,  the Hapke albedos are consistently the highest, apart from regions of strong absorption by ice at $\lambda > 1~\mu\mathrm{m}$. $g_{\rm delta\text{-}Eddington}$ is higher than $g_{\rm Shkuratov}$, so the delta-Eddington albedo should be lower than the Shkuratov albedo (Fig. \ref{fig4}). However, since $(1-\bar{\omega})_{\rm delta\text{-}Eddington} < (1-\bar{\omega})_{\rm Shkuratov}$, the delta-Eddington albedo is usually intermediate between the albedos estimated by Hapke and Shkuratov models.  

The actual asymmetry parameter of the firn and ice is likely to be intermediate between  $g_{\rm Shkuratov}$ and $g_{\rm delta\text{-}Eddington}$. This can be explained by examining the delta-Eddington model we use, which assumes that scattering within firn and glacier ice takes place by \textit{spherical} bubbles of air trapped within ice. However, as Figures \ref{fig1}b and \ref{fig1}c show, the bubbles are far from spherical. \citet{dadic2013effects} pointed this out, noting that nonspherical particles scatter less toward the forward direction, and more to the side \citep[Figure 7 of][]{neshyba2003hexagonal}; \citep[Figure 4 of][]{grenfell2005hollow}. Thus, \citet{dadic2013effects} were able to obtain similar fits to our best-fit grain radii simulations by using the measured SSA, and instead multiplying the delta-Eddington model's asymmetry parameter by factors ranging from 0.78 to 0.94. Based on their calculations, the true asymmetry parameter of the firn and ice likely ranges from 0.67 to 0.82.  

Typically, a reduction in $g$ would result in higher albedos, all else being equal, as mentioned above. But when accounting for specular reflection, the albedo reduces overall, except at $\lambda = 1.5~\mu\mathrm{m}$ (see Section \ref{sec3.2}). Thus, when accounting for specular reflection \textit{and} scattering by nonspherical bubbles, $g$  must be reduced to raise the albedo and match the measurements.  

For the Hapke model to match the albedos for $\lambda < 1~\mu\mathrm{m}$ better, $g_{\rm Hapke}$ would have to be increased, but at the cost of worsening the fit at longer wavelengths. However, the fact that the Hapke model can fit the data somewhat reasonably (errors $<0.12$) using $g_{\rm Hapke}=0$ (Figs. \ref{fig2}a-c) shows that inferring ice grain radii from the Hapke model can give good spectral fits using unrealistic physical parameters. Single-scattering theory cannot produce $g=0$ for ice at solar wavelengths for any grain radius \citep[Figure 2 and Figure 3 of][]{mishchenko1994asymmetry} or shape \citep{dang2016effect}. $g_{\rm Hapke}$ can,  however, be estimated if the phase function of the scatterers is available, although measuring the phase function can be challenging \citep{hudson2006spectral, cull2010compositions}. If $g_{\rm Hapke}$ is estimated using measurements of the phase function, the quantities for $\gamma$ and $\bar{\omega}$ (Eq.~2) are replaced by $\gamma^*$ and $\bar{\omega}^*$, respectively, in the Hapke model, using Equations 11.21a and 11.21b of \citet{hapke2012theory}. Regardless, increasing $g_{\rm Hapke}$ to reduce the model albedos by increasing the mean photon path lengths would be inconsistent with assuming spherical particles in the Hapke model, because the model mean photon path lengths cannot exceed those produced by spherical particles \citep[Figure 2 of][]{Shkuratov2005}.  

Although no information about the sphericity of the scatterers is provided explicitly when we run the Shkuratov model, values for $g_{\rm Shkuratov}$ are quite close to the true asymmetry parameter derived for firn and ice by \citet{dadic2013effects}. Thus, the Shkuratov model fits the measurements best for $\lambda < 1~\mu\mathrm{m}$ (Figs. \ref{fig2}a-c). We note that the Shkuratov model was formulated using the average path length between two internal reflections required as input \citep[Equation 4 of][]{shkuratov1999modeling}, which we have interpreted as being equivalent to an effective particle diameter  \citep{Poulet2002}  so that the model results can be interpreted in terms of measurable quantities such as SSA. Thus, it may be possible for the Shkuratov model results to improve if the average path length between two internal reflections was measured and input directly into the model.  Note that if non-spherical particles are assumed for the Shkuratov model, the reduction in mean photon path length from approximately  $0.9 \times 2r$ to $0.2 \times 2r$ \citep{Shkuratov2005} would result in higher model albedos (all else being equal), leading to a greater discrepancy between the Shkuratov model albedos and the measured albedos.  

The third reason for the discrepancies between the modeled results and the measurements is whether the models can account for specular reflection at the surface. Of the three models, only the delta-Eddington model we use was designed to account for this effect, which becomes important for lake ice, firn, and glacier ice. Thus, when specular reflection is included, the delta-Eddington model outperforms the Hapke and Shkuratov models for $\lambda > 1~\mu\mathrm{m}$ (Figs. \ref{fig3}a-c).  

Based on our quantitative comparison, we recommend using the delta-Eddington model for predicting the albedo and inferring the grain radius of ices across the solar system because it (a) generally produces the least error in predicting the albedo, and (b) it is based on realistic physical parameters. It also accounts for specular reflection that may occur for snow that has metamorphosed into firn and bubbly glacier ice (e.g., the slabs of CO$_2$ ice that form in the martian polar regions).  We note that the Shkuratov and Hapke models provide better, or similar matches to the grain radii while failing to match the measured albedos because those models do not account for: (1) the increased absorption within dense ice \citep[which may be accounted for using radiative transfer methods ranging widely in complexity;][]{bohren1983radiative, mullen1988theory, vaisanen2020scattering}, and (2) specular reflection at the surface of firn and ice. All three models do not account for the nonsphericity of bubbles within ice. Future work would benefit from conducting additional, well-constrained measurements of field albedo, SSA, impurity, and density of snow, firn, and ice on Earth. They could be used to help further test these models quantitatively before they are applied to ices across the solar system to infer the physical properties of the surface and predict the albedo.

\section*{Data Availability}
The measured downward solar spectrum and the spectral albedo observations of firn and glacier ice made by \cite{dadic2013effects} are available at \url{https://digital.lib.washington.edu/researchworks/items/dda68540-9276-4dc8-8d06-0eb751a7e048}. Results from the radiative transfer models are available at: \url{https://zenodo.org/records/15136427}.

\section*{Acknowledgments}
A part of this work was carried out at the Jet Propulsion Laboratory, California Institute of Technology, under a contract with the National Aeronautics and Space Administration. We would like to thank Editor Gianrico Filacchione, two anonymous reviewers, Steve Warren for very helpful advice and edits, Dale Winebrenner for helpful conversations, and Ružica Dadić for sharing the full spectral-resolution measurements.
\section*{Declaration of competing interest}
The authors declare no competing interests

\bibliographystyle{elsarticle-harv} 
\bibliography{ref}
\end{document}